\documentclass[11pt,twoside]{article}

\usepackage{asp2021}

\aspSuppressVolSlug
\resetcounters

\begin{document}

\title{Astronomy Open Science Competence Centre in Europe}

\author{Marco~Molinaro,$^1$ Mark~Allen,$^2$  Joachim Wambsganns,$^3$ Enrique~Solano,$^4$ Baptiste~Cecconi,$^5$ Markus~Demleitner,$^3$ Andr\'{e}~Schaaff,$^2$ Hendrik~Heinl,$^1$ and Sara~Bertocco$^1$}
\affil{$^1$INAF - Astronomical Observatory of Trieste, Trieste, Italy; \email{marco.molinaro@inaf.it}}
\affil{$^2$CNRS / CDS, Strasbourg, France}
\affil{$^3$University of Heidelberg, Heidelberg, Germany}
\affil{$^4$INTA CSIC, Madrid, Spain}
\affil{$^5$Paris Observatory, Paris, France}

\paperauthor{Marco~Molinaro}{marco.molinaro@inaf.it}{0000-0001-5028-6041}{INAF}{Osservatorio Astronomico di Trieste}{Trieste}{}{}{Italy}
\paperauthor{Mark~Allen}{mark.allen@astro.unistra.fr}{}{CNRS}{CDS}{Strasbourg}{}{}{France}
\paperauthor{Joachim~Wambsganns}{jkw@ari.uni-heidelberg.de}{}{Universit\"{a}t Heidelberg}{Astronomisches Rechen-Institut}{Heidelberg}{}{}{Germany}
\paperauthor{Enrique~Solano}{esm@cab.inta-csic.es>}{}{INTA}{CSIC}{Madrid}{}{}{Spain}
\paperauthor{Baptiste~Cecconi}{baptiste.cecconi@obspm.fr}{}{}{Observatoire de Paris}{Paris}{}{}{France}
\paperauthor{Markus~Demleitner}{msdemlei@ari.uni-heidelberg.de}{}{Universit\"{a}t Heidelberg}{Astronomisches Rechen-Institut}{Heidelberg}{}{}{Germany}
\paperauthor{Andr\'{e}~Schaff}{andre.schaaff@astro.unistra.fr}{}{CNRS}{CDS}{Strasbourg}{}{}{France}
\paperauthor{Hendrik~Heinl}{hendrik.heinl@inaf.it}{}{INAF}{Osservatorio Astronomico di Trieste}{Trieste}{}{}{Italy}
\paperauthor{Sara~Bertocco}{sara.bertocco@inaf.it}{}{INAF}{Osservatorio Astronomico di Trieste}{Trieste}{}{}{Italy}



\begin{abstract}
The Astronomy Open Science Competence Centre Pilot (Astro-CC) is an ESCAPE-cluster related project meant to enable the
astronomy research communities to accelerate their use of Open Science by supporting the implementation of FAIR principles.

The Astro-CC project aims at expanding the use of Virtual Observatory standards by astronomy-focused ESFRIs, RIs, and
data-producing projects of all scales, enabling the astronomy research communities to accelerate their use of Open Science
by supporting the implementation of FAIR principles. It will run community events engaging experts in astronomical data \&
service interoperability to prepare and define the scope of a Community Competence Center.

The project will support the community development of the Virtual Observatory interoperability framework and its
integration into EOSC, building on the progress made in the ESCAPE Science Cluster project. It aims at contributing to the
vision of EOSC as a federation, providing feedback on the practical implementation of Open Science to the EOSC
governance.
\end{abstract}



\section{The Project}
The Horizon Europe Open Science Clusters' Action for Reasearch \& Society
(OSCARS\footnote{\url{https://oscars-project.eu/}}) has a goal to foster the uptake of Open Science in Europe,
consolidating past achievements of the thematic Science Clusters into FAIR data service and working practices.

To support the research communities to take up Open Science and foster the involvement of scientists in EOSC,
OSCARS provided Open Calls for cascading grants to support researchers in uptaking FAIR principles.

The Astronomy Open Science Competence Centre Pilot project
(Astro-CC\footnote{\url{https://www.oscars-project.eu/projects/astro-cc-astronomy-open-science-competence-centre-pilot}})
is one of the projects funded by the first OSCARS' Open Call, and it is related to the ESCAPE Science Cluster (including
astronomy, nuclear and particle physics), continuation of the Horizon 2020 ESCAPE project.

Astro-CC will help the astronomy research communities in the uptake of Open Science and supporting the
implementation of FAIR principles. The project will support the community development of the Virtual Observatory
interoperability framework and its integration into EOSC. The Virtual Observatory open standards are the basis
for making astronomy data FAIR.

The complexity of managing astronomical data across various platforms and ensuring that it adheres to the FAIR principles, 
requires a community approach. The community also needs training and improved access to tools that support data 
interoperability and sharing across multiple RIs.

Astro-CC addresses this challenge by running Community Competence Centre events, providing training on the
implementation of interoperable services, the development of the underlying Virtual Observatory data sharing framework
and its integration in EOSC, and exploitation of the system for scientific research. These events (better described in
Sec.~\ref{sec:events}) will specifically target different facets of the astronomy communities:
\begin{itemize}
	\item Data Provider Forum event;
	\item Technology Forum events;
	\item Scientific Training events.
\end{itemize}

The project will engage national Research Infrastructures (RIs) and expand into interdisciplinary fields, such as
planetary science and heliophysics. These efforts aim to promote interoperability across diverse astronomical
domains. Astro-CC activities are aimed at extending the participation beyond the ESCAPE Science Cluster 
astronomy/astro-particle fields to include the wider astronomy communities of planetary science, heliophysics
and space weather.

The community events of this project will engage the European experts in astronomical data/
service interoperability to prepare and define the scope of a Competence Centre that will serve
the extended range of astronomical communities.

Astro-CC leverages the opportunities of the OSCARS program to advance the connection of disciplinary interoperability
frameworks to EOSC, contributing to the vision of EOSC as a federation, and providing feedback on the practical
implementation of Open Science to the EOSC governance.

\section{The Events}
\label{sec:events}

As mentioned, the Astro-CC events will be specifically targeted to engage different facets of the
astronomy communities.
\begin{description}
	\item [A Data Provider Forum] event: for Research Infrastructures and data-intensive projects to
share best practices for the implementation of standards to provide services for FAIR data.
	\item [Technology Forum] events: for the developers of astronomical software/services; to discuss
community open standards and approaches to FAIR in the astronomy (in its largest definition)
research domain.
	\item [Scientific Training] events: for astrophysics Ph.D students and early career researchers; to learn using
interoperable tools and services as well as to gain skills for Open Science publication and research in astronomy.
These training events leverage operational services, software and training tutorials that have been on-boarded 
in EOSC.
\end{description}

At the time of writing a couple of events took already place:
\begin{itemize}
	\item the 1st Technology Forum\footnote{1st Tech Forum website: \url{https://indico.in2p3.fr/event/36318/}},
held in Trieste in the second week of October 2025;
	\item the 1st Scientific Training\footnote{1st Training Event website:
\url{https://indico.in2p3.fr/event/36738/}}, held in Madrid in the first week of December 2025;
\end{itemize}
while the Data Providers Forum\footnote{Data Providers Forum website: \url{https://indico.in2p3.fr/event/37439/}}
is under organisation and will happen in Heidelberg at the end of March 2026.

A small report of the 1st Tech Forum is available in this proceeding (see Sec.~\ref{sec:itf}).
For the training and the providers' forum it's possible to find further information on the events' websites (linked
here above from the footnotes).

\section{The 1st Astro-CC Tech Forum}
\label{sec:itf}

The 1st Technology Forum event of Astro-CC took place at the Basovizza Observing Station of the INAF-OATs (Trieste, Italy)
with remote participation available for talks and (best effort) hands-on ``hack-a-thons''.
It was attended by 30 participants reporting in talks and/or discussing or doing hands-on work on:
\begin{itemize}
	\item VO standards' updates;
	\item Planetary data publishing;
	\item EPRV spectra;
	\item FAIRness, semantics and metrics;
	\item AI relationship to VO standards;
	\item VO relationship to other platforms;
	\item Contributing the PyVO effort.
\end{itemize}
The format balanced 4 plenary sessions for talks with 10 hack-a-thon slots; all the slots where 90' long.
An intro session and a wrap up enclosed the nearly 3 days of continuous work.

The forum provided a valuable venue for the community to connect and discuss the topics listed above, clarifyng open points about existing standards, exemplifying existing technologies, discussing community requirements towards new standardisation needs, and allowed for a place to consider what it means a thematic node in EOSC for the community of astrophysics.

\section{Next Steps}

The Astro-CC project is moving towards the end of its first year of activities. At the end of the first 12 months, 3 out of
the proposed 5 events will be closed and a first report could be done. The remaining part of the project will see the 2nd 
Technology Forum and the 2nd Scientific Training take place, so far under definition in terms of location and timeframe.
One goal of the project itself will be clarifying what would be the goal, benefits, and efforts required in implementing and
maintaining a thematic node for astrophysics (at large) within the Federation of the EOSC Nodes.

\acknowledgements The OSCARS project has received funding from the European Commission's Horizon Europe
Research and Innovation programme under grant agreement No.101129751. The Astro-CC project has been funded
through the first Cascading Grants call of the OSCARS project.



\end{document}